# STUDY OF ORGANIC MONOLAYER MODIFIED METAL OXIDE SEMICONDUCTOR DEVICES FOR HIGH TEMPERATURE APPLICATIONS

Rosilin George, U. Satheesh, J. Cyril Robinson Azariah, D. Devaprakasam*
NEMS/MEMS and Nanolithography Lab,
Department of Nanosciences and Technology,
Karunya University, Coimbatore-641 114
Email:devaprakasam@karunya.edu

**Abstract:** We report fabrication and characteristics of an organic monolayer based Metal Oxide Semiconductor (MOS) device. In place of $SiO_2$ oxide layer in the MOS configuration, we used 1H, 1H, 2H, 2H- perfluorooctyl trichlorosilane (FOTS) self-assembled monolayer as a substitution. The MOS device was fabricated by simple steps like sputter deposition and dip coating method. The device was heat treated to different temperatures to understand its performance and efficiency for high temperature application. The MOS device was heated to 150° C, 350°C and 550°C and the energy band gap was found to be varied in the order 2.5 eV, 3.0 eV and 3.4eV respectively. For non-heated sample, the energy band gap is 3.4 eV. The results shows that the parameters like charge mobility (μ), energy band gap, and resistance were found to be decreased after the heat treatment. The change in the energy band gap due to heat treatment has significantly influenced the I-V and the impedance characteristics. We observed that the MOS device started to conduct between 1V to 3V, further the device conduct till 20V. Impedance analysis show that device heated to 350 °C shows the low impedance but the impedance starts to increase for further heating up to 550°C. Using Multi Dielectric Energy Band Diagram Program (MEBDP) we studied the MOS structure and C-V characteristics and temperature dependent behavior of the devices from 100K to 600K. Our experimental work and simulation studies confirm that the FOTS SAM substituted MOS device could be used for high temperature applications. The experimental observations are well supported by the simulation results. This study shows that the FOTS organic monolayer are promising substitute for $SiO_2$ oxide layer in the MOS and MOSFET which could be used high temperature applications.

## INTRODUCTION

In the field of electronics, inorganic semiconducting materials like Germanium and Silicon took over the important role after the invention of transistors. But in the recent days, organic semiconductors show better characteristics and it becomes new trend to use them in the electronics industry. Study of organic semiconductors is not new but was triggered again by the discovery of Photoluminescence studies related to charge carrier transport and optical excitation [1, 2, and 3].

Organic Semiconductor devices [4] have a wide range of applications like Organic LED passive matrix display [5]. Organic filed effect transistor [6], sensors, and memory devices. The ease in electronic tuning and other processing properties increases the demand for flexible organic devices. The achievement in organic/inorganic transistor includes compatibility with p and n channel organic/inorganic semiconductors; very low gate leakage current and very good thermal and chemical stability, enhanced capacitance versus voltage characteristics and efficient fabrication methods.

Self-assembling monolayers are 2D material of great interest. The properties of the novel device are improved mainly due to the self-assembling property of FOTS since ordering and orientation have a great influence in device properties. Tuning of work function will be possible due to the presence of self-assembling monolayer. Studies have proved that the semiconductor morphology has been altered when deposited SAM modified Au on it [7, 8]. Self-assembled monolayers (SAMs) of organic molecules have good insulating properties and large capacitance values [9].

Here we report the study of a Metal Organic Semiconductor device and its improvement in properties is conformed after heat treatment with the help of various characterization techniques. For characterization, instruments like JFC-1600 Auto Fine coater, NI-PXI workstation, Solartron SI-1260 impedance analyzer, Tubular furnace, and Photoluminescence Spectroscopy is used.

## METHODOLOGY

### 1.1 Materials and Methods

FOTS is an organic silane molecule, which are sandwiched between the silicon wafer and the gold layer. FOTS is an organic self-assembled monolayer and it is a promising agent for anti-stiction coatings [10]. The structural representation of FOTS is shown in Figure 1.

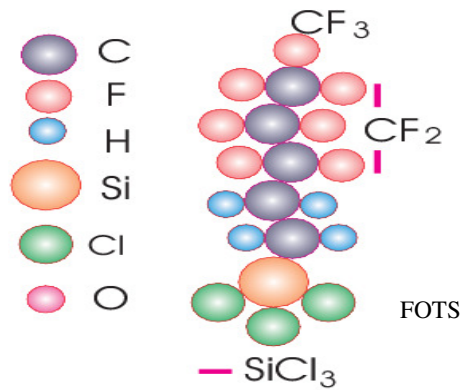

Figure 1.Shows the schematic representation of a FOTS molecule.

Boron doped p-type silicon wafer is used as the substrate material and gold act as a metal layer. FOTS and gold is grafted on to the p-Si wafer.

**1.2 Fabrication of Metal Organic Semiconductor Device**

Boron doped p-type Silicon wafer is used as the substrate material [11, 12]. The wafer is cleaned using acetone. 1mM of FOTS solution is prepared using isooctane as the solvent. FOTS layer was formed over the p-Si substrate material by dip coating method, where the substrate is immersed in the solution for an hour, which formed a Langmuir Blodgett (LB) Film [13, 14, and 15] over the Silicon wafer. A Gold (Au) metal layer is formed over the coated FOTS SAM layer by sputter coating method.

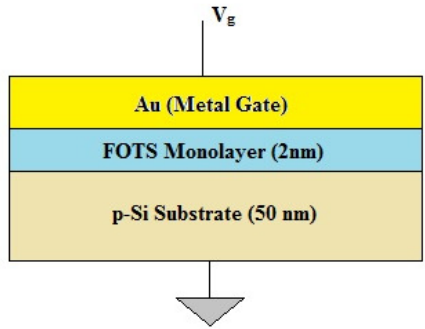

Figure 2.Shows the schematic of the fabricated MOS device.

Connections are made on the fabricated device using silver wire and silver conductive paste. Figure 2 shows the fabricated MOS device. The fabricated device is then heat treated using tubular furnace. The device is heated at 200°C, 400°C and 600°C for one hour each and the characteristics were analyzed for each sample. Heat treatment contributed noticeable variation in characteristics for the device. The prepared sample is then. The properties of the FOTS SAM and Au coated on p-Si substrate after heating for different temperatures like 150ºC, 350ºC and 550ºC are investigated in this study.

**1.3 Heat treatment studies**

The prepared samples were taken for the heat treatment studies and heated in a tubular furnace for different temperatures like $150^0$C, $350^0$C and $550^0$C for one hour each and the characteristics of each sample were analysed. Heat treatment studies contributed noticeable variations in the characteristics of the device.

**1.4 I-V and impedance characterization method**

I-V characteristic is measured using NI-PXI work station, which includes synchronisation buses and other hardware features required for the I-V measurements. Solartron SI-1260 impedance analyser is used to measure capacitance, resistance and inductance for desired frequencies. It can be used with DC bias and AC signal at different sweep and frequency conditions. JFC-1600 Auto Fine Coater is used to coat the gold layer over the FOTS SAM layer. The device consists of an argon gas purifying system which permits efficient sputtering. Energy band gap is calculated with the help of photoluminescence spectroscopy.

The principle involves absorption and emission of photons (electromagnetic radiation).

**RESULTS AND DISCUSSIONS**

**3.1 Photoluminescence Studies**

The heat treatment studies produced variations in the band gap energy of each device which affected the electron flow rate and conduction through the device. The band gap of the stack is calculated using equation (4):

$$E = h\upsilon \quad (1)$$

Where E is energy; H is Plank's constant; $\upsilon$ is frequency.

$$c = \lambda\upsilon \quad (2)$$

Where, c is speed of light ($3\times 10^8$ m/sec); $\lambda$ is wavelength; $\upsilon$ is frequency.

From equation (2),

$$\upsilon = c/\lambda \quad (3)$$

Combining equations (1) & (3), we get

$$E = hc/\lambda \quad (4)$$

E = 6.626 x $10^{-34}$ J·s x 3 x $10^8$ m/sec/ ($\lambda$)

E = 1.988 x $10^{-25}$ J·m/$\lambda$ m E

Figure 3 shows the combined photoluminescence graph for the non-heat treated and heat treated samples. Band gap energy was calculated with respect to the highest peaks corresponding to each graph. Band gap energy of 2.7eV, 3.1eV and 3.4eV were obtained for the samples heated treated 150˚C, 350˚and 550˚C respectively. Energy band gap was increasing with the increase in heat treatment temperatures.

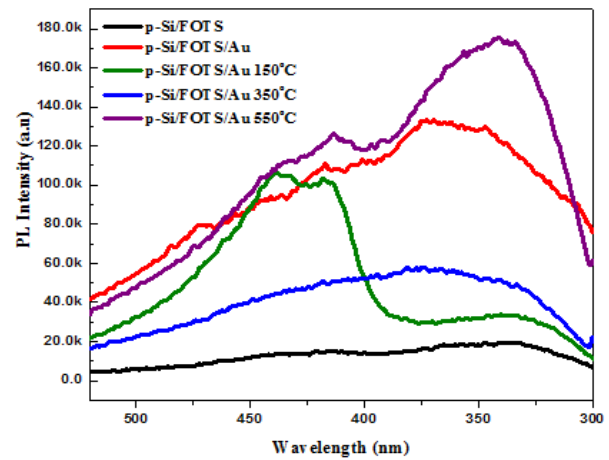

Figure 3 Combined Photoluminescence Graph shows the FOTS and Au coated p-Si wafer for non-heat treated and heat treated samples

Table 1shows the band gap energy and corresponding wavelengths for Au/FOTS/p-Si MOS configuration.

| Sample/ Temperature | Wavelength (λ) nm | Band gap Energy(E) eV |
|---|---|---|
| Si/FOTS/Au | 347 | 3.4 |
| Si/FOTS/Au 200˚C | 439 | 2.7 |
| Si/FOTS/Au 400˚C | 382 | 3.1 |
| Si/FOTS/Au 600˚C | 342 | 3.4 |

The table 1 gives a clear idea about the changes takes place in the band gap energy of the MOS device for the prepared and heat treated samples.

### 1.5 I-V Measurements

The fabricated device is connected to the NI-PXI instrument using silver wire connection leads for I-V measurements. We heat treated three samples at 150ºC, 350ºC and 550ºC are taken for this study. After the heat treatment, we characterized the samples; the I-V characteristic of the heat treated sample is shown in the Figure 4. With the rise in voltage between 2V to 3 V, the MOS device starts to conduct. Even at 10V the device is conducting before breakdown occurs. This proves that the device is suitable for high temperature applications.

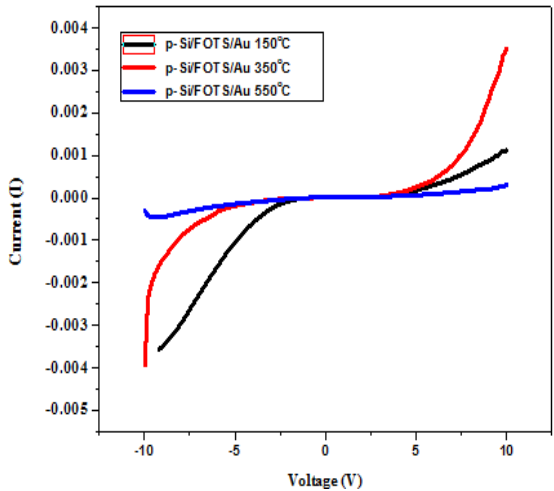

Figure 4. I-V curve shows the annealed samples of FOTS and Au coated p-Si wafer.

### IMPEDANCE SPECTROSCOPY

Figure 8 shows the combined impedance curve for the three heat treated samples at 150ºC, 350ºC and 550ºC. The Cole-Cole plot is obtained for all the heat treated samples with different values as shown in Figure 5.

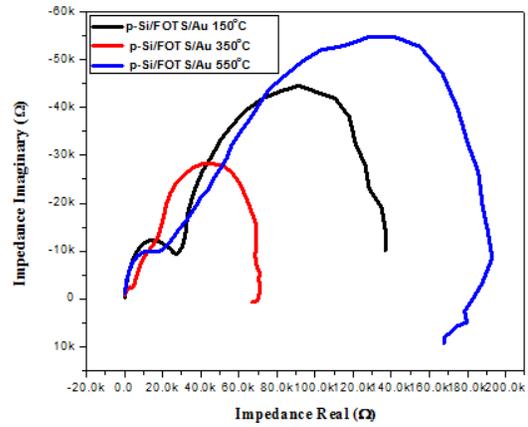

Figure, 5. Shows the Cole-Cole Plot of the heat treated samples of FOTS and Au coated p-Si wafer.

After the heat treatment studies, the resistance of the device was found to be decreased, but when heating is continued above a particular temperature, the resistance started to increase again due to increase in band gap. From the Cole-Cole plot, it is observed that up to 350 ºC, the impedance starts decreasing gradually but further heating starts to increase the impedance up to 550 ºC. Figure 6 shows the impedance versus frequency curve for the heat treated samples of Au/FOTS SAM coated p-Si wafer.

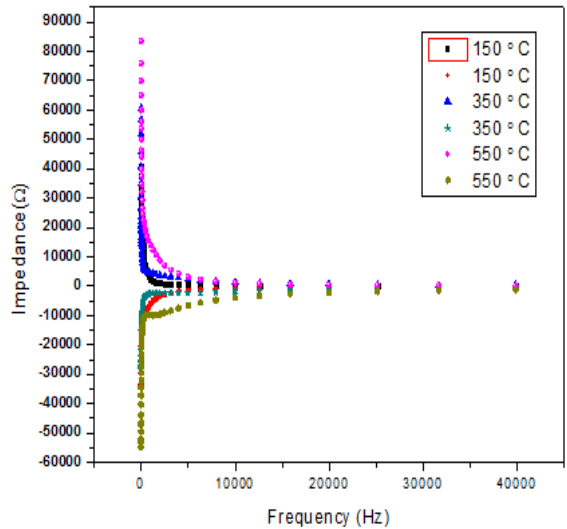

Figure 6 shows the real and imaginary Impedance versus frequency

### 1.6 Simulation Results

For the Au/FOTS/p-Si MOS configuration, the following simulations are generated using the MEBDP simulation software. It includes the energy band diagram at the flat-band condition, stack capacitance versus applied voltage and gate charge versus applied voltage ($Q_g$-V Characteristics).

Au was used as the top metal layer with work function as 4.8 with the thickness of 4 nm, FOTS material with dielectric constant 4 with the thickness of 2nm, electron affinity 2.39 (eV) and band gap 5(eV) . We have used p-type silicon wafer with dielectric constant 11.7 and electron affinity 4.05 (eV) and the dopant concentration greater than intrinsic carrier concentration.

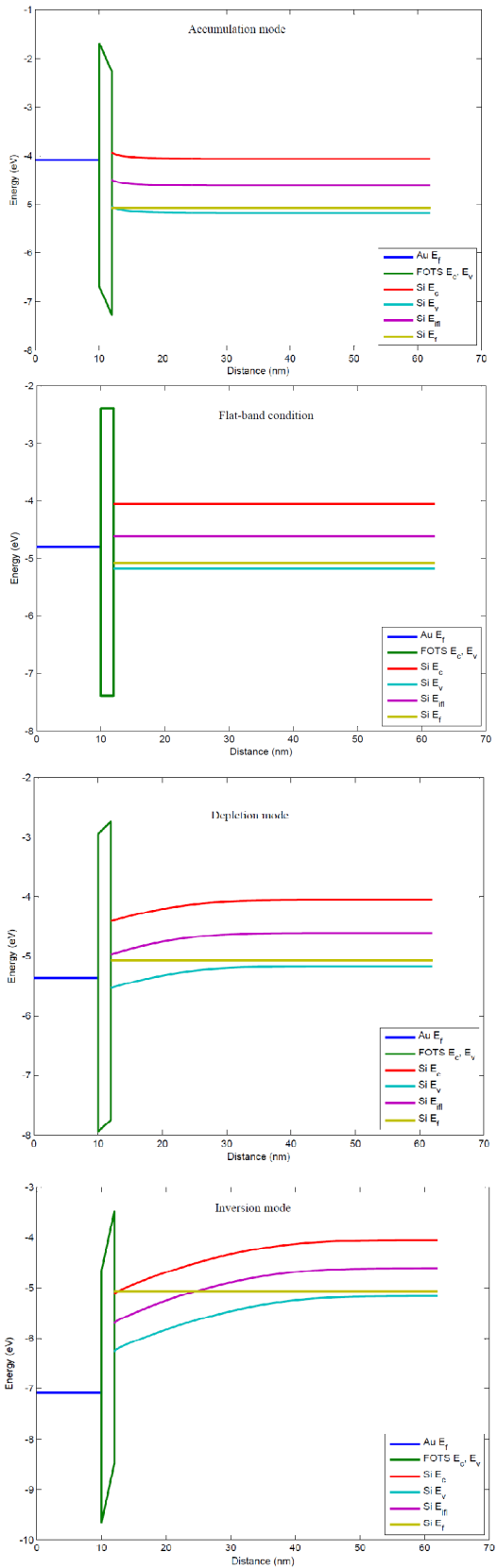

flat-band condition, a negative bias of -0.28V is applied. The dynamic characteristic of the device is studied using the simulation plot between the gate-charge versus the applied voltage characterization as shown in Figure 8.

From the MOS C-V curve, accumulation region appears on introducing high negative bias to the gate with respect to the bulk or body. The flat-band voltage is -0.28V, beyond which the device reaches the accumulation mode. Depletion requires applying the small positive potential (say between 0.1V to 0.5V) to the gate with respect to the bulk or body.

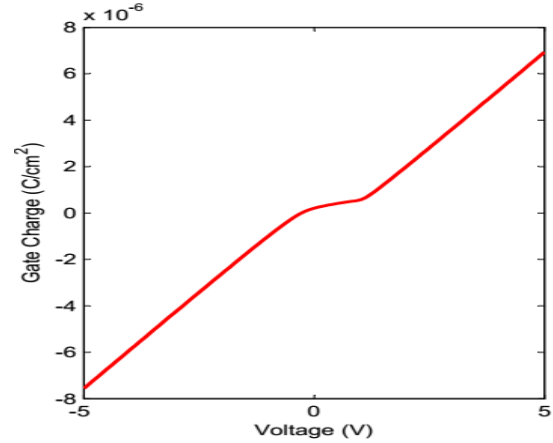

Figure 8 shows the simulated plot between the gate charge versu the voltage ($Q_g$-V characteristics) of the Au/FOTS/p-Si configuration.

The majority carriers, which are holes being positively charged carriers pushed away from the surface under the effect of this gate potential. The depletion region is filled carriers which are negatively charged acceptor ions.

On increasing the positive bias (say between 0.5V to 1.5V) of the gate with respect to the body leads to weak inversion. As a result, electron-hole pairs are generated in the depletion region near the surface. The electric field is then created by the gate potential, separates electron-hole pair before they can recombine and pulling the negatively charged electrons to the surface. Hence the surface looks like weakly n-type now. The onset of strong inversion occurs when applying a very high positive potential (say above +2V) to the gate with respect to the body.

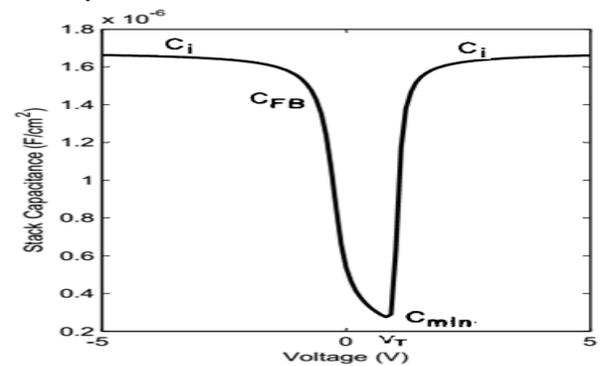

Figure 7. shows the energy band diagram of Au/FOTS/p-Si configuration at the flat-band condition.

The energy band diagram simulation was carried at flat-band condition with no charge in the bulk region. To obtain the

Figure 9 shows stack capacitances versus the voltage (C-V characteristics) of the Au/FOTS/p-Si Configuration.

The characteristics of the stack was analysed with the help of MEBDP at different temperatures. In this study, we observed that, threshold voltage of the stack decreases with increase in temperature.

Table 2 shows the stack capacitance ($C_{stack}$), Capacitance of FOTS ($C_{FOTS}$), Capacitance of Silicon ($C_{Si}$) and voltage drop of FOTS and Silicon wafer during accumulation, flat band, and depletion and inversion modes of the stack.

| Modes | $C_{stack}$ F/cm² x10⁻⁶ | $C_{FOTS}$ F/cm² x10⁻⁶ | $C_{si}$ F/cm² x10⁻⁶ | FOTS Voltage drop (V) | Si Voltage drop (V) |
|---|---|---|---|---|---|
| Accumulation | 1.714 | 1.771 | 53.32 | -1.545 | -0.175 |
| Flat-band | 1.043 | 1.771 | 2.535 | 0 | 0 |
| Depletion | 0.284 | 1.771 | 0.3338 | 0.277 | 0.753 |
| Inversion | 1.691 | 1.771 | 37.44 | 1.186 | 1.094 |

Table 3 shows the flat-band voltage, effective oxide thickness and threshold voltage of the simulated device.

| Temperature (K) | Flatband voltage (V) | EOT (nm) | Threshold voltage (V) |
|---|---|---|---|
| 200 | -0.347 | 1.950 | 1.033 |
| 400 | -0.206 | 1.950 | 0.901 |
| 600 | -0.043 | 1.950 | 0.751 |

From Table 3, flat-band voltage $V_{FB}$, threshold voltage $V_{TH}$ ($C_{stack}$) and EOT of the simulated MOS structure at a temperature of 150ºC, 350ºC and 550ºC. It is noted that the flat-band voltage and the threshold voltage of the simulated device decrease.

Table 4. Shows the various capacitance values obtained from the C-V graph.

| Insulator Capacitance '$C_i$' (F/cm²) | 1.65x10⁻⁶ |
|---|---|
| Minimum Capacitance '$C_{min}$' (F/cm²) | 0.3x10⁻⁶ |
| Flatband Capacitance '$C_{FB}$' (F/cm²) | 1.4x10⁻⁶ |
| Threshold Voltage '$V_{TH}$' (V) | 1 |

The C-V characteristic of the stack device obtained from the simulator which is illustrated in Figure 9. The insulator capacitance ($C_i$), minimum capacitance ($C_{min}$), Flat band capacitance ($C_{FB}$) and threshold voltage ($V_{th}$) calculated is given in Table 4.

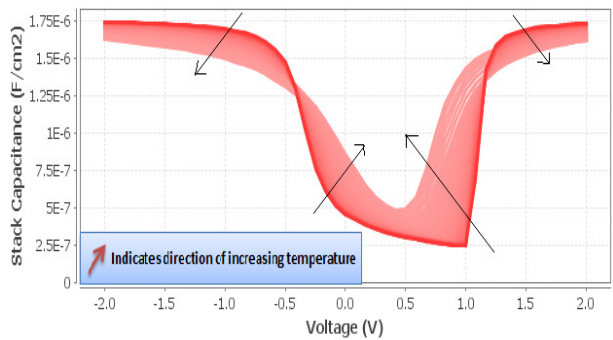

Figure. 10. shows the simulated stack capacitance with respect to temperature ranging from 100K-600K.

From the figure 10, it is observed that the stack capacitance increases on increasing the temperature whereby the threshold voltage decreases. Hence the conduction through the device is also increased.

## SUMMARY

An organic monolayer modified metal oxide semiconductor device was fabricated using simple techniques like dip coating method and sputter deposition. The fabricated samples were heat treated to different temperatures using tubular furnace. Heat treatment produced noticeable changes in band gap energy, conductivity and resistivity of the fabricated devices. It is confirmed that the parameters of the device influenced by the heat treatment at a particular range of temperature. IV characteristics showed that the conduction in the devices started between thresholds of 2V to 3V. It is observed that the devices are conducting even up to 20V without causing any damage or breakdown to the device. From the impedance it is observed that up to 350 ºC, the impedance starts decreasing gradually but further heating starts to increase the impedance up to 550 ºC. The changes in the impedance are due to order- disorder behaviour of FOTS SAM which was caused by heat treatment. Using MEBDP we simulated MOS structure and estimated C-V characteristics and temperature dependent behavior of the devices from 100K to 600K. Our experimental work and simulation studies confirm that the FOTS SAM substituted MOS device could be used for high temperature applications.

## ACKNOWLEDGEMENT

We thank DST-Nanomission, Government of India and Karunya University for providing the financial support to carry out the research. We thank the department of Nanosciences and technology for the help and support to this research. We are grateful to Mr. Alfred Kirubaraj, Assistant Professor, ECE, Karunya University for the useful suggestion and help.